\documentclass{elsart}
\usepackage{apjfonts}
\usepackage{epsfig}
\usepackage{natbib}
\usepackage{amssymb}
\usepackage{floatflt}

\def\aap{{A\&A}}

\def\amin{$^\prime$}
\def\aj{{AJ}}
\def\apj{{ApJ}}

\def\apjs{{ApJS}}
\def\asec{$^{\prime\prime}$}
\def\mnras{{MNRAS}}
\def\nat{{Nature}}
\def\pasp{{PASP}}

\def\gtrsim{\mathrel{\hbox{\rlap{\hbox{\lower4pt\hbox{$\sim$}}}\hbox{\raise2pt\hbox{$>$}}}}}

\newcommand{\halpha}{H\ensuremath{\alpha}}

\newcommand{\hbeta}{H\ensuremath{\beta}}

\newcommand{\hst}{\emph{HST}}

\newcommand{\kms}{km~s\ensuremath{^{-1}}}

\newcommand{\mbh}{\ensuremath{M_\mathrm{BH}}}

\newcommand{\msigma}{\ensuremath{M_{\mathrm{BH}}-\sigmastar}}
\newcommand{\msun}{\ensuremath{M_{\odot}}}

\newcommand{\sigmastar}{\ensuremath{\sigma_{\ast}}}

\newcommand{\spitzer}{\emph{Spitzer}}

\def\lax{{$\mathrel{\hbox{\rlap{\hbox{\lower4pt\hbox{$\sim$}}}\hbox{$<$}}}$}}
\def\gax{{$\mathrel{\hbox{\rlap{\hbox{\lower4pt\hbox{$\sim$}}}\hbox{$>$}}}$}}

\begin{document}

\begin{frontmatter}

\title{The Smallest AGN Host Galaxies}

\author{Jenny E. Greene}
\address{Harvard-Smithsonian Center for Astrophysics, 60 Garden St., 
Cambridge, MA 02138}

\author{Aaron J. Barth}
\address{Department of Physics and Astronomy, University of California
at Irvine, 4129 Frederick Reines Hall, Irvine, CA 92697-4575}

\and

\author{Luis C. Ho}
\address{The Observatories of the Carnegie Institution of Washington,
813 Santa Barbara St., Pasadena, CA 91101}

\begin{abstract}

We describe our efforts to study dwarf galaxies with active nuclei, whose
black holes, with masses $\leq 10^6$~\msun, provide the best current
observational constraints on the mass distribution of primordial seed
black holes.  Although these low-mass galaxies do not necessarily contain
classical bulges, Barth, Greene, \& Ho (2005) show that their stellar
velocity dispersions and black hole masses obey the same relation as more 
massive systems.  In order to characterize the
properties of the dwarf hosts without the glare of the active nucleus, we
have compiled a complementary sample of narrow-line active galaxies
with low-mass hosts.  The host galaxy properties, both their structures
and stellar populations, are consistent with the general properties of
low-mass, blue galaxies from Sloan.  The black holes in these galaxies
are probably radiating close to their Eddington limits,
suggesting we may have found Type 2 analogues of narrow-line
Seyfert 1 galaxies.

\end{abstract}

\end{frontmatter}

\section{Motivation}

It is now widely accepted that supermassive BHs play an integral role
in the formation of galaxies (Ho 2004), as manifested in the tight
correlations between black hole (BH) mass and both bulge luminosity
(e.g.,~Marconi \& Hunt 2004) and stellar velocity dispersion
(the \msigma\ relation; Gebhardt \etal\ 2000; Ferrarese \& Merritt
2000; Tremaine \etal\ 2002).  Although the exact nature of the
connection remains 

%%%%%%%%%%%%%%%%%%%%%%%%%%%%%%%%%%%%%%%%%%%%%%%%%%%%%%%%%%%%%%%%%%
\hskip 1.2cm
\vskip -0.1in
\vbox{
\hbox{
\hskip +0.12in
\psfig{file=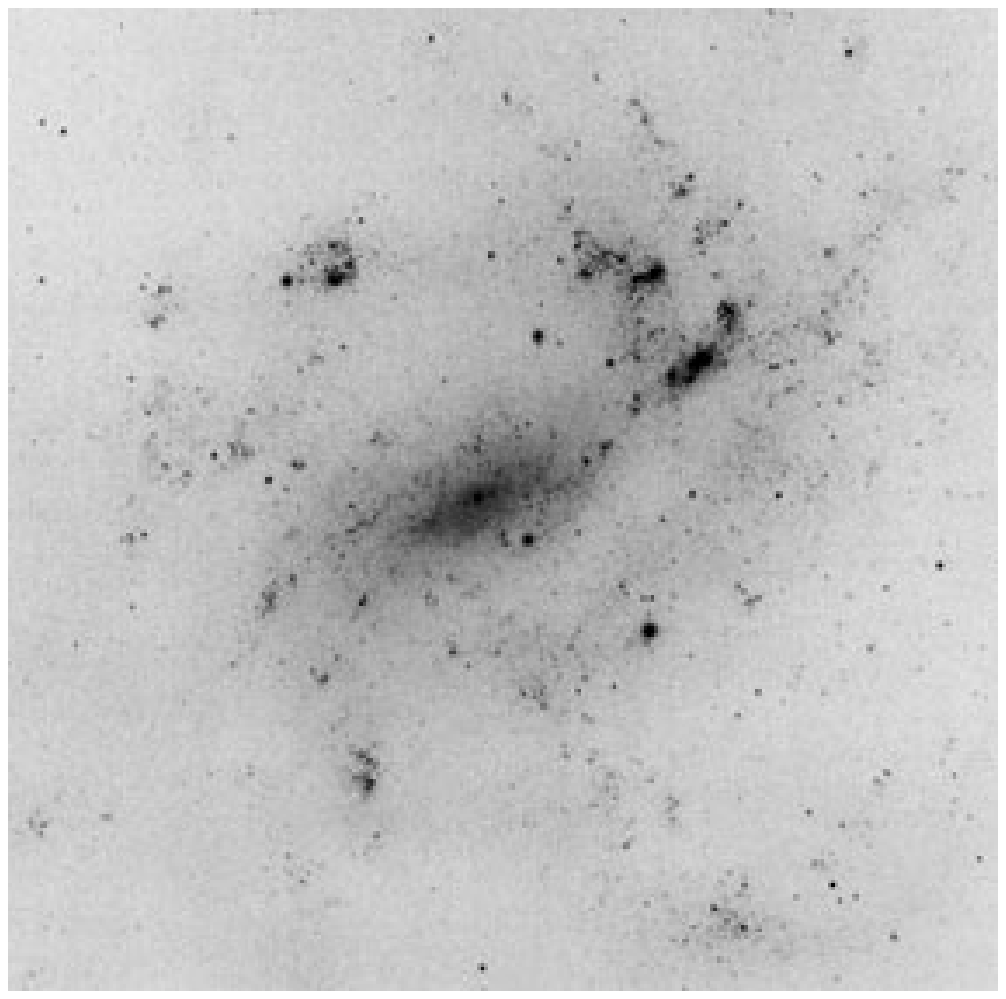,height=2.4truein,angle=0}
\hskip +0.16in
\psfig{file=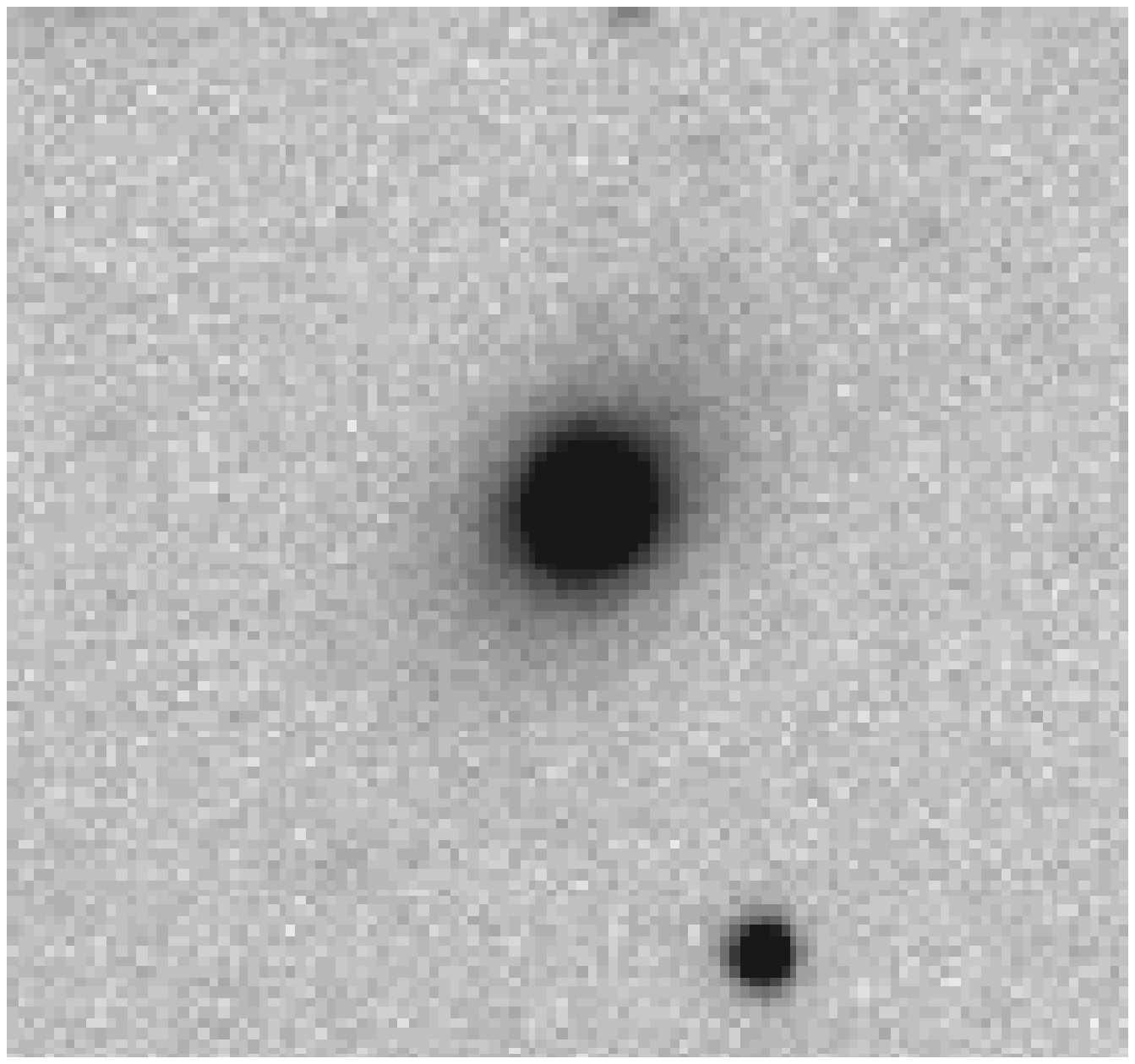,height=2.4truein,angle=0}
}
}
\vskip -0.1in
{\small
\noindent{\it Figure 1} --- 
Two examples of AGNs in late-type galaxies.  {\it Left:}\ 
optical image of NGC 4395, $\sim$15\amin\ (17 kpc) on a side.  {\it Right:}\
$R$-band image of POX 52, $\sim$25\asec\ (11 kpc) on a side.  }
%%%%%%%%%%%%%%%%%%%%%%%%%%%%%%%%%%%%%%%%%%%%%%%%%%%%%%%%%%%%%%%%%%

unclear, a variety of theoretical models suggest that feedback from
the active galactic nucleus (AGN) is responsible for truncating both
star formation and continued BH growth (Silk \& Rees 1998; Kauffmann
\& Haehnelt 2000; Di Matteo \etal\ 2004).  AGN feedback may be
responsible for more than just the \msigma\ relation itself; it may be
drive other galaxy properties, including the
observed color and mass bimodality (e.g.,~Strateva \etal\ 2001;
Kauffmann \etal\ 2003a; Springel \etal\ 2004).  In fact, the vast
majority of local AGNs are found in massive, bulge-dominated galaxies
(e.g.,~Ho \etal\ 1997b; Kauffmann \etal\ 2003b).  While BHs in dwarf
galaxies are apparently rare, they provide a unique probe of the
relation between galaxies and BHs.  We expect such BHs to have low
masses ($10^4-10^6$~\msun), and thus to provide an important
probe of the properties of primordial seed BHs.  Current dynamical
techniques do not have sufficient resolution to detect BHs of such small
mass outside the Local Group, while studies of the Scd galaxy M33
(Gebhardt \etal\ 2001) and the dwarf elliptical NGC 205 (Valluri
\etal\ 2005) suggest that these Local Group dwarf galaxies do not host
supermassive BHs.  In the absence of direct dynamical evidence, we
must rely on radiative signatures (AGN activity) to indicate the
presence of a BH for more distant systems.

\section{The Search for Intermediate-mass BHs}

The best-studied examples of intermediate-mass BHs in active galaxies
are NGC 4395 (e.g.,~Filippenko \& Sargent 1989), with Hubble type Sdm,
and POX 52 (Barth \etal\ 2004), a dwarf elliptical (Fig. 1).  While
their nuclear optical spectra and broader spectral energy
distributions (e.g.,~Moran \etal\ 1999; Barth \etal\ in prep) are much
like other AGNs, the galaxies are much smaller.  At least in the case
of NGC 4395, there is no bulge component.
And yet, their nuclear stellar velocity
dispersions and BH masses are roughly consistent
with the low-mass extrapolation 

%%%%%%%%%%%%%%%%%%%%%%%%%%%%%%%%%%%%%%%%%%%%%%%%%%%%%%%%%%%%%%%%%%
\vskip-2cm
\hskip 1.cm
\psfig{file=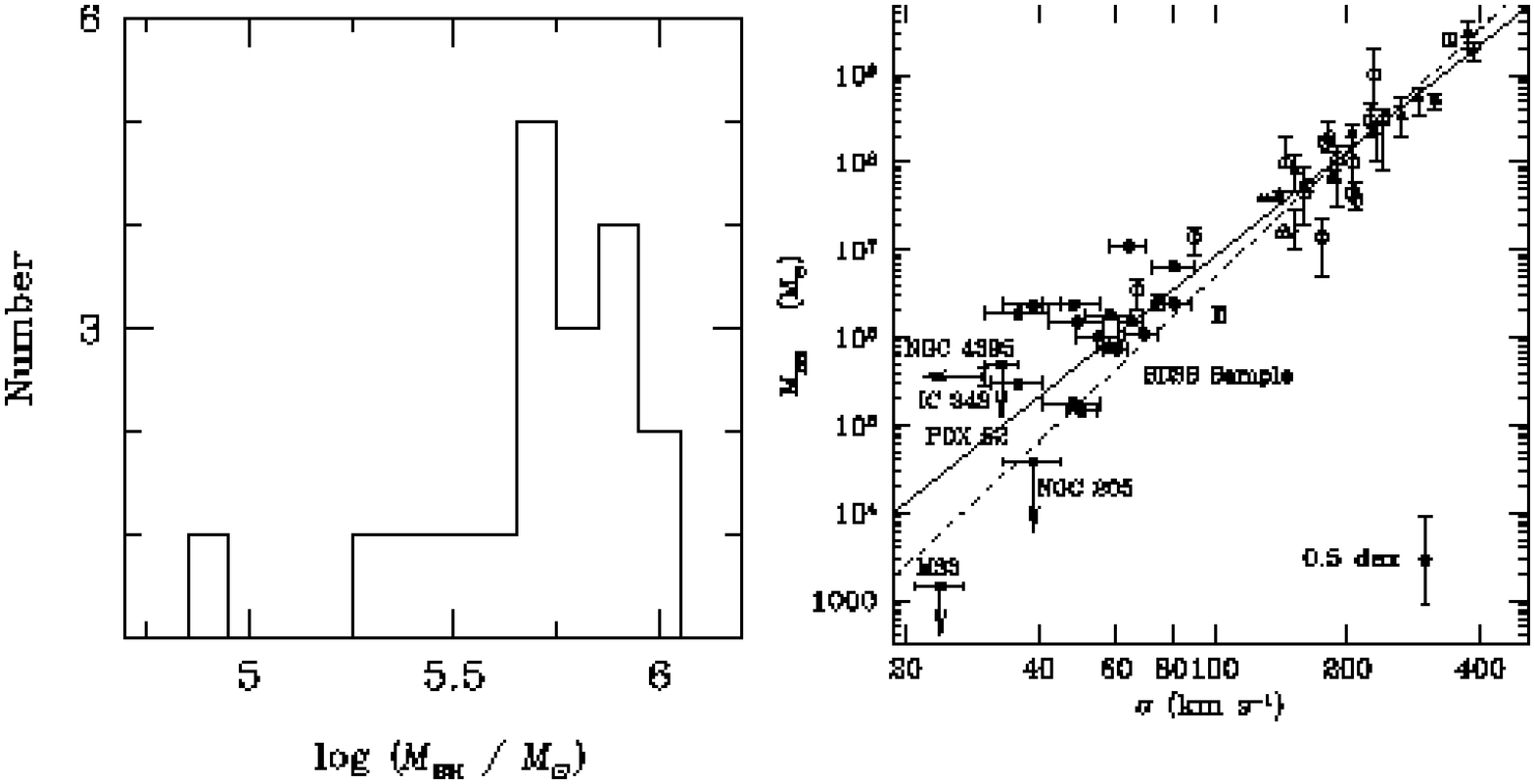,width=2.3truein,angle=-90}
\vskip -0.1in
{\small
\noindent{\it Figure 2} --- {\it Left:}\ Distribution of BH masses for the 
new sample of 19 SDSS broad-line AGNs with candidate intermediate-mass BHs 
from Greene \& Ho (2004).   {\it Right:}\ The \msigma\ relation for central 
BHs in galaxy nuclei (Barth, Greene, \& Ho 2005).  The open circles and filled
triangles come from Tremaine et al. (2002).  Upper limits to the BH masses in 
the nearby, inactive galaxies M33 (Gebhardt et~al. 2001), NGC 205
(Valluri \etal\ 2005) and IC 342 
(B{\" o}ker et~al. 1999), as well as mass estimates for NGC 4395
(Peterson \etal\ 2005) and POX 52 (Barth \etal\ 2004), 
are shown.  The filled circles are 17 of the 19 SDSS objects.  
The solid and dashed lines represent the \msigma\ relation 
derived by Tremaine et al. (2002) and Merritt \& Ferrarese (2001), 
respectively.
}
\vskip -0.1in
%%%%%%%%%%%%%%%%%%%%%%%%%%%%%%%%%%%%%%%%%%%%%%%%%%%%%%%%%%%%%%%%%%

of the \msigma\ relation (Filippenko \& Ho 2003; Barth \etal\ 2004;
Fig. 2; we also inclue the BH in the globular cluster G1; Gebhardt
\etal\ 2005).  While intriguing, it is hard to draw definite
conclusions from three objects alone.  We need a larger sample.  For
this reason, Greene \& Ho (2004) conducted a systematic search for
intermediate-mass BHs, finding 19 in the First Data Release of the
Sloan Digital Sky Survey.  After selecting {\it all} broad-line AGNs
with $z < 0.35$, to ensure that \halpha\ was in the bandpass, they
used the linewidth-luminosity mass scaling relations\footnote{These
``virial'' mass estimates employ a velocity dispersion from a
broad-line width (in this case \halpha; Greene \& Ho 2005) and a
broad-line region size from the radius-luminosity relation defined for
reverberation-mapped AGNs.}  of of Kaspi et al. (2000) to define a
sample with \mbh\ $< 10^6$ \msun\ (Fig. 2).  Although the selection
was based on \mbh, the derived host luminosities (dashed lines in
Fig. 3) are $\sim 1$ mag below $L^*$; these are small galaxies.
Subsequent stellar velocity dispersion measurements from ESI on Keck
revealed that remarkably, the objects are consistent with the low-mass
extrapolation of the \msigma\ relation (Fig. 2; Barth \etal\ 2005).
What does \sigmastar\ represent for these galaxies?  In the case of
NGC 4395, it is the velocity dispersion of the central nuclear star
cluster.  For the remaining objects, only \hst\ has the resolution
required to determine the photometric properties of the stellar
nucleus (Barth \etal\ in prep; Greene \etal\ in prep).  Determining
the stellar ages and metallicities of the host galaxies is complicated
by emission from the broad-line AGN.  In order to study dwarf host
galaxies in more detail, and complete the demographics of
intermediate-mass AGNs, we search for narrow-line 

%%%%%%%%%%%%%%%%%%%%%%%%%%%%%%%%%%%%%%%%%%%%%%%%%%%%%%%%%%%%%%%%%%
\hskip 1.cm
\psfig{file=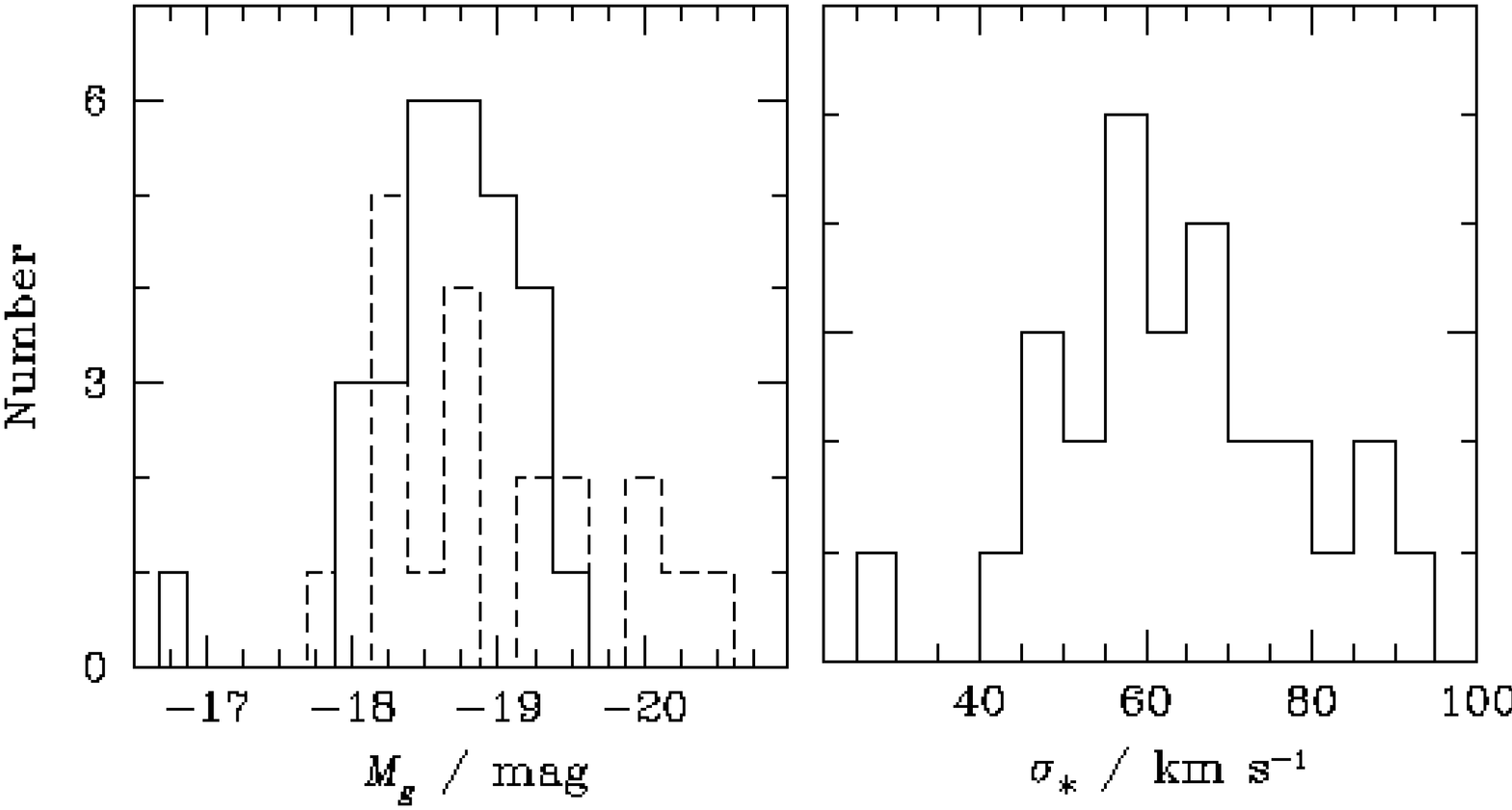,width=2.5in,keepaspectratio=true,angle=-90}
\vskip -4.mm
{\small
\noindent{\it Figure 3} --- {\it Left}: Host galaxy $g$-band absolute
magnitudes for the broad-line (dotted) and narrow-line (solid)
samples.  {\it Right}: Distribution of stellar velocity dispersions
for the narrow-line sample, as measured from the Keck spectra (Barth,
Greene, \& Ho, in prep). }
%%%%%%%%%%%%%%%%%%%%%%%%%%%%%%%%%%%%%%%%%%%%%%%%%%%%%%%%%%%%%%%%%%
\noindent
AGNs with low-mass BHs.

\section{The Narrow-line Sample}

In the absence of broad lines, we cannot repeat the search technique
employed by Greene \& Ho (2004) because these objects have no virial
BH mass estimates.  Instead, since Barth \etal\ (2005) show that the
\msigma\ relation extends to \mbh\ $< 10^6$ \msun, we simply search
for low-mass galaxies with narrow-line AGN spectra.  We use the
catalog of 33589 narrow-line AGNs from Kauffmann \etal\ (2003b), who
have kindly made available \sigmastar\ measurements for the entire
sample (Brinchmann \etal\ 2004).  Of all objects with \sigmastar\ $<
70$ \kms\ (unresolved in Sloan) and $M_g < -19.5$ (see Fig. 3), we
have obtained follow-up spectroscopy for 29 objects with ESI on Keck.
Using a direct pixel-fitting method described in Barth \etal\ (2002),
we find a range in stellar velocity dispersions of 25 to 95 \kms.  The
stellar masses from Kauffmann \etal\ (2003a) are $M_{\ast} \sim
10^9$~\msun, and colors and structural parameters from the Sloan
pipeline are typical of the blue, low-mass galaxies in that catalog.  
The Sloan images (Fig. 4) already show us that we have a mix
of compact and disk-like systems, but \hst\ imaging will be required
to determine the true structural compositions of these galaxies.  The
most extreme object in our sample, SDSS J1109+6123, with a stellar
mass of $M_{\ast} = 10^{8.1}$~\msun\ and an absolute magnitude of
$M_g=-16.8$ mag, is the {\it smallest} known AGN host galaxy.  The ESI
spectra will allow us to model the stellar ages and metallicities in
more detail, as shown in the composite spectrum (Fig. 5).  We will
then investigate how the stellar populations, structure and
environments for this sample compare to inactive galaxies of similar
mass.

%%%%%%%%%%%%%%%%%%%%%%%%%%%%%%%%%%%%%%%%%%%%%%%%%%%%%%%%%%%%%%%%%%
\vskip -1cm
\hskip 2.5cm
\psfig{file=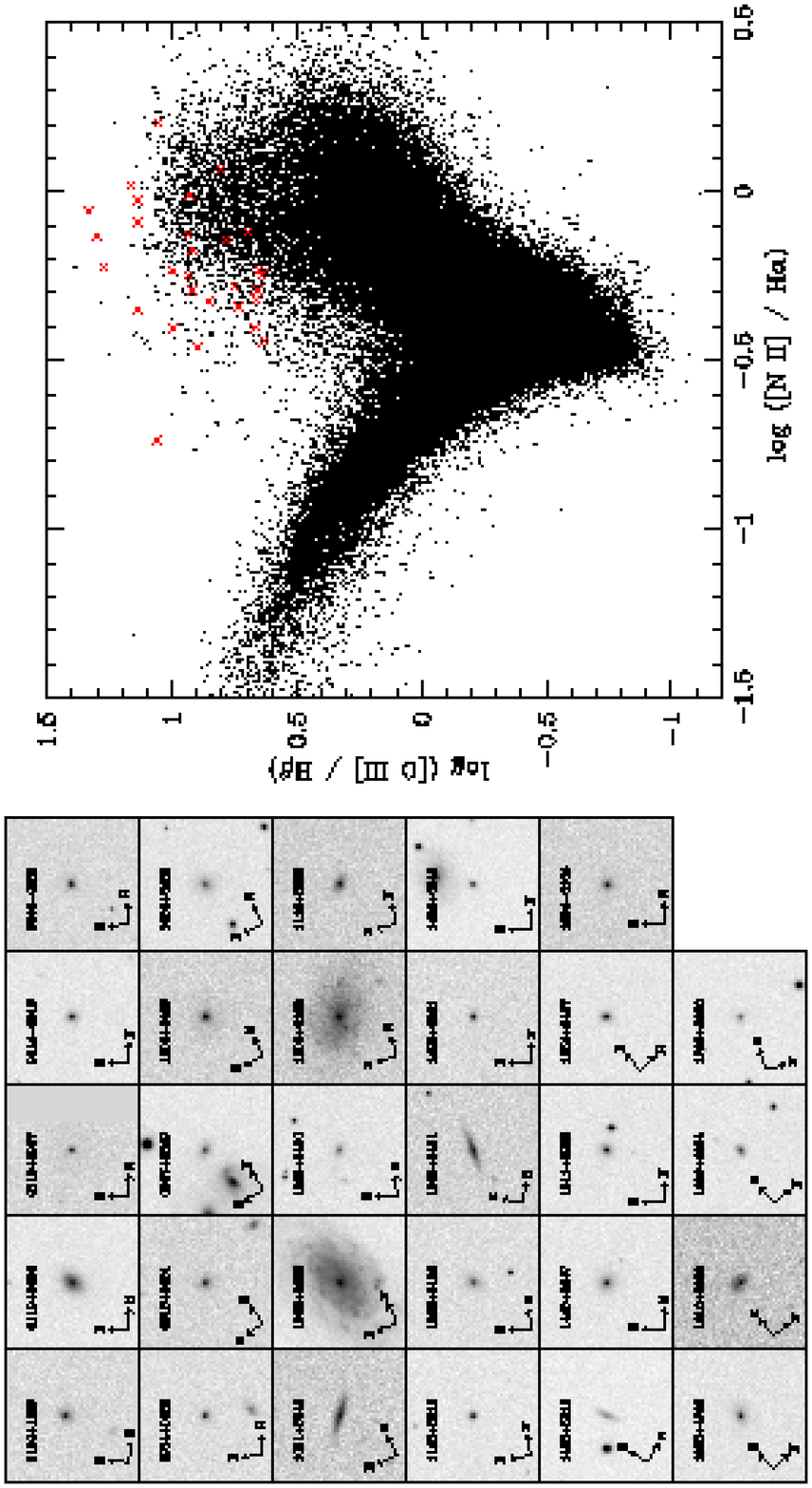,width=2.3in,keepaspectratio=true,angle=-90}
%\makebox(0,0)[t]{ \bf Please see high resolution version of the paper
%  for this figure.}
\vskip -1.mm
{\small
\noindent{\it Figure 4} --- 
{\it Left}: Sloan images for the narrow-line sample.
{\it Right}: The diagnostic diagram of [O {\footnotesize
III}]/\hbeta\ vs. [N {\footnotesize II}]/\halpha\ for the Kauffmann \etal\
(2003a) sample.  The values for Type 2 AGNs from the ESI spectra 
(Barth, Greene, \& Ho, in prep) are shown as red crosses.  The line
ratios clearly fall far from the stellar locus.  For reference, the
standard defining values of Seyfert line ratios 
are [N {\footnotesize II}]/\halpha\ 
\gax\ 0.6 and [O {\footnotesize III}]/\hbeta\ \gax\ 3 (e.g.,~Ho \etal\ 1997a).
}
%%%%%%%%%%%%%%%%%%%%%%%%%%%%%%%%%%%%%%%%%%%%%%%%%%%%%%%%%%%%%%%%%%

While the host galaxies are extremely atypical, the emission-line
ratios for the sample clearly demonstrate that these are truly
accretion-powered AGNs (Fig. 4).  The inferred range of BH masses,
assuming the Tremaine \etal\ (2002) fit to the \msigma\ relation, are
$10^{4.5} <$\mbh$< 10^{7}$ \msun.  If we convert the [O {\footnotesize
III}] luminosities into bolometric luminosity using the conversion of
Heckman \etal\ (2004), we find that, much like the Greene \& Ho (2004)
objects, the narrow-line objects are radiating close to their
Eddington limits.  Perhaps we have found Type 2 analogues of
so-called narrow-line Seyfert 1 galaxies (NLS1s; Osterbrock \& Pogge
1985).  In this regard, {\it XMM-Newton} and \spitzer\ observations
will be crucial to quantify their absorbing columns and their
bolometric luminosities.  Greene, Ho, \& Ulvestad (2006) found that
the broad-line sample is uniformly radio-quiet, much like the NLS1
population in general.  Since only one of the narrow-line objects is
detected by FIRST, we suspect this sample will be similar, but deeper
radio images at higher resolution are required to test this.

%%%%%%%%%%%%%%%%%%%%%%%%%%%%%%%%%%%%%%%%%%%%%%%%%%%%%%%%%%%%%%%%%%
\vskip -0.5cm
\hskip 1.5cm
\psfig{file=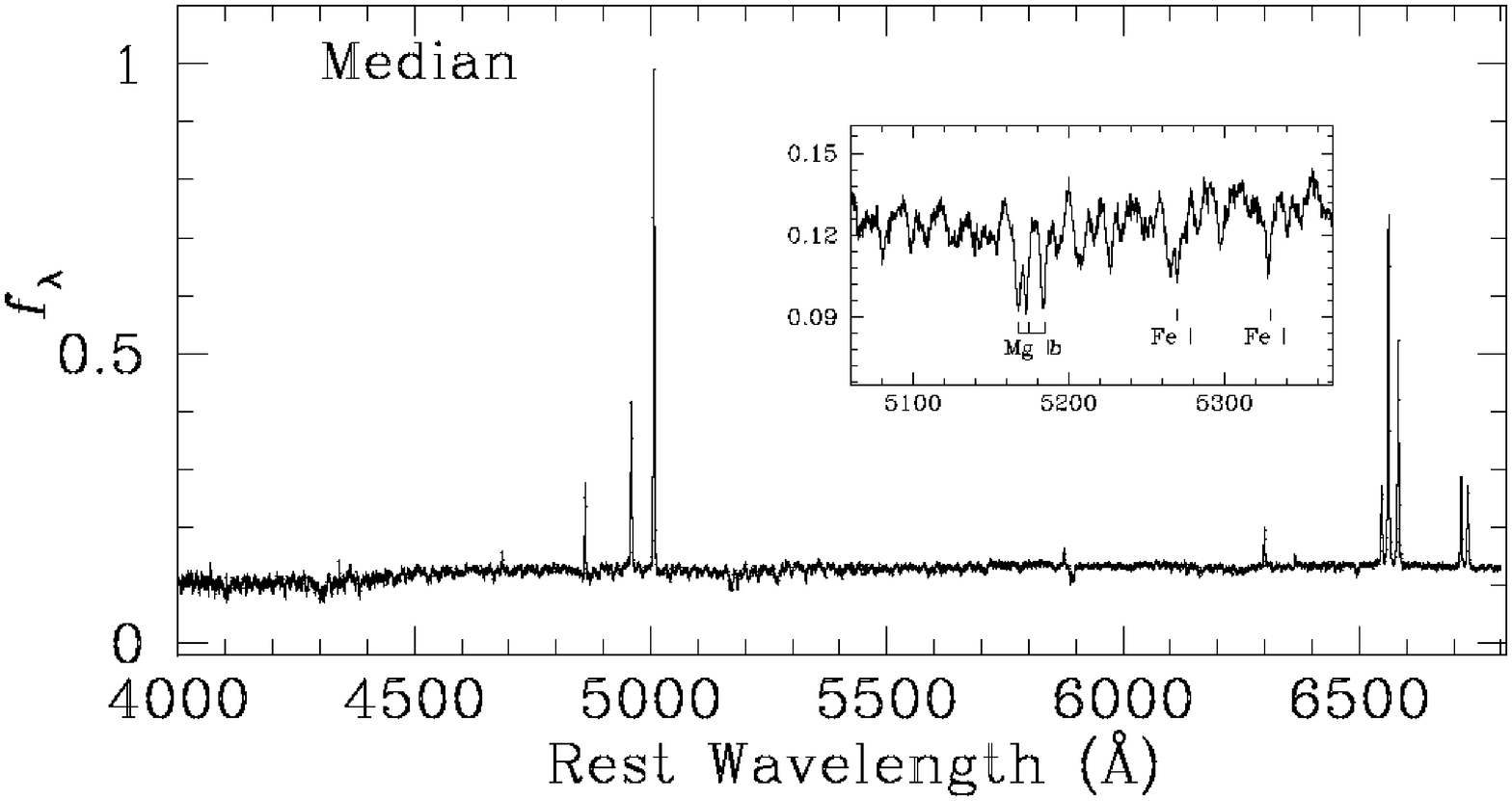,width=1.9in,keepaspectratio=true,angle=-90}
\vskip -4.mm
{\small
\noindent{\it Figure 5} --- 
The composite ESI spectrum for the sample of 29 narrow-line active
galaxies. The flux is normalized arbitrarily.
The continuum is rich in stellar absorption features.  
}
%%%%%%%%%%%%%%%%%%%%%%%%%%%%%%%%%%%%%%%%%%%%%%%%%%%%%%%%%%%%%%%%%%

\end{document}